\documentclass[11pt,onecolumn]{IEEEtran} 

\usepackage{graphicx}
\usepackage{amssymb}
\usepackage{amsmath}
\usepackage{amssymb}
\usepackage{array}
\usepackage{xy}

\def\BibTeX{{\rm B\kern-.05em{\sc i\kern-.025em b}\kern-.08em
    T\kern-.1667em\lower.7ex\hbox{E}\kern-.125emX}}

\newtheorem{thm}{Theorem}[section]

\newtheorem{lem}[thm]{Lemma}

\begin{document}

\title{Low Latency Wireless Ad-Hoc Networking: Power and Bandwidth Challenges and a Hierarchical Solution} 

\author{Nima Sarshar, Behnam A. Rezaei and Vwani P. Roychowdhury\\ $\{nima,behnam,vwani\}@ee.ucla.edi$}

\markboth{}
{} 

\maketitle \setlength{\baselineskip}{6.2mm}

\begin{abstract}
This paper is concerned with the scaling of the number of hops in
a large scale wireless ad-hoc network (WANET), a quantity we call
network latency. A large network latency affects all aspects of
data communication in a WANET, including an increase in delay,
packet loss, required processing power and memory. We consider
network management and data routing challenges in WANETs with
scalable network latency.  On the physical side, reducing network
latency imposes a significantly higher power and bandwidth demand
on nodes, as is reflected in a set of new bounds. On the protocol
front, designing distributed routing protocols
 that can guarantee the delivery of data packets within scalable
 number of hops is a challenging task. To solve this, we introduce multi-resolution randomized hierarchy (MRRH), a novel
power and bandwidth efficient WANET protocol with scalable network
latency. MRRH uses a randomized algorithm for building and
maintaining a random hierarchical network topology, which together
with the proposed routing algorithm can guarantee efficient delivery
of data packets in the wireless network. For a network of size $N$,
MRRH can provide an average latency of only $O(\log^{3} N )$. The
power and bandwidth consumption of MRRH are shown to be
\emph{nearly} optimal for the latency it provides. Therefore, MRRH,
is a provably efficient candidate for truly large scale wireless
ad-hoc networking.

\end{abstract}


\section{Introduction} Efficient data delivery in wireless ad-hoc
networks (WANET's) is a challenging task due to the lack of  any
global coordination of the nodes in the network. Network management
protocols, as well as routing algorithms have to work in an
autonomous manner, and yet overall, the network should be able to
efficiently route traffic from any source to any destination. Most
routing strategies are also interested in minimizing the average
power required for network operation, which requires the use of
short-distance communication. The ease of routing and power
considerations are, perhaps, the primary reasons why most known
WANET routing protocols resort to simple, Nearest-Neighbor
Communication (NNC) strategies and their variants, for routing.

Nearest neighbor communication, however, can not scale properly as
the system size grows, and WANETs of increasingly larger sizes
become realizable. In its simplest form, for a network of $N$
nodes, the NNC strategy will use $\sim \sqrt{N}$ relay nodes for
an average communication. We call {\em the number of hops a WANET
uses for an average communication}, its {\bf\em network latency}
(NL). The name is chosen because this form of latency is induced
by the network layer, and is a reflection of the underlying
routing algorithm \footnote{Note that network latency, unlike
delay, is  independent of the communication rate. Delay, on the
other hand, is usually defined as the reciprocal of communication
rate (or throughput), and is a measure of the amount of time
required to communicate a given volume of data between two
points.}. The average network latency of $\Theta (\sqrt{N})$ could
significantly impact the system performance of NNC based routing
schemes, as summarized
below. \\
(1) Increasing the NL will increase the communication delay due to
buffering at relay nodes. In steady state, one expects the
average {\em delay to be linearly proportional to the NL}.\\
(2) As NL increases, the probability of end-to-end packet loss
increases. This may require excessive number of packet retransmissions, which in turn would
increase average power consumption. Such tradeoffs have recently received considerable attention (see e.g., \cite{hc1,hc2,hc3,hc4}).\\
 (3) An increase in NL
 induces computational and memory overhead on a larger
 number of nodes in the network. In a nearest neighbor
 communication scheme, e.g., a node has to process and route one packet
 out of every $\Theta(\sqrt{N})$ packets transmitted in the
 network which directly translates into a need for more  processing power.\\
Therefore, NNC is most suited for small to medium size applications.

\bigskip
 {\em New
routing schemes} with {\em significantly better NL performance}
have to be devised for the WANETs, if they want to scale to truly
large sizes; perhaps with millions of participants.  Designing a
large-scale low-NL WANET, however, brings about many physical as
well as algorithmic challenges that are considered next. This
paper focuses on  wireless ad-hoc networking schemes with low
network latency, and the effect of reducing the number of
communication hops on critical performance measures of the
networks, including power, bandwidth and throughput.

\subsection{Low Network-Latency WANET's: Challenges}
Distributed operation of a WANET becomes a significant challenge when NL has
to be low, i.e., packets have to be delivered within only a small
number of hops.  The
reason is that a node has to expect data packets from other nodes that
are far away in the network. This is unlike NNC models where a node has to manage its communication
with only a limited number of spatially close by nodes. As described next, scheduling routes, multiple access control (e.g.,
time and frequency division multiplexing), as well as, code
management become particularly nontrivial in  a low NL WANET.

To be specific, lets consider the hypothetical and extreme scenario
of single hop or direct communication. Suppose also that the
bandwidth and power requirements are not issues, and thus, to get
the shortest possible NL  (i.e., one), any source node directly
communicates with its destination node. To eliminate interference,
each communication pair can be assigned an independent
non-interfering channel with enough bandwidth. Assume that at any
round, every node is communicating with another, randomly chosen,
node. For a network of size $N$, therefore, a total of at least
$N/2$ communication channels have to be set up for each round of
communication; something that might be prohibitive even for
moderately large size $N$. Now, in the next round, each node will
decide to change its communication partner. How would a given node
notify its destination node about the channel in which the
communication has to take place? How does the network, in a
decentralized way, schedule and assign channels to communication
pairs?  This task seems impossible, unless there is one dedicated
channel for every pair of nodes (a total of $N(N-1)/2$ channels),
which is certainly impractical. Even then, this requires every node
to have global information about every other node in the network.

None of the above problems would have been an issue in an NNC strategy.
There, a node is only responsible for
receiving and relaying packets to a constant number of close
neighbors. Multiple access control (MAC) can be performed by local
negotiations, and one of the many local routing algorithms could have
been adopted, a few of which will be reviewed later in this section.

Since single-hop communication is impractical in ad-hoc situation
({\em even if} power and bandwidth are no constraints), we then
ask the question, whether distributed protocols exist for
efficient joint multiple access control and data routing (similar
to NNC), while providing a low NL (similar to direct
communication)? This paper proposes one such solution,
Multi-Resolution Randomized Hierarchy (MRRH), based on randomized
data structure concepts.

Apart from MAC and routing challenges, low NL communication is
demanding in terms of bandwidth and power requirements.
Reducing the number of hops  requires the use of longer
communication links, which in turn requires more power. Increasing
the power for a communication link, on the other hand, will
increase the interference on other nodes in the network, which
calls for an increased bandwidth. Any proposed low NL
networking protocol has to be evaluated with respect to these
tradeoffs. To quantify this, {\em we derive a new set of fundamental
tradeoffs} among  NL, node density,
throughput, average power, number of channels and total bandwidth
of any WANET. These bounds quantify the tradeoffs between various
performance measures of a WANET. Our proposed scheme, MRRH, is then
 shown to nearly satisfy these bounds; in other words, {\em power
and bandwidth requirement of MRRH is nearly optimal given the
 NL it provides}.

\subsection{Performance Bounds and Proposed Algorithm}
Our performance bounds relate NL to different metrics of a WANET
and are derived in Section \ref{sec:bounds}. As shown in Theorem
\ref{prop:power}, for a direct line-of-sight power dissipation,
the average power consumption of any WANT with NL of $L$ should
scale as $\Omega(N/L^2)$, where $N$ is the network size,
regardless of the bandwidth. Similarly, Theorem \ref{prop:bnd}
states that the total bandwidth required by any WANET with NL of
$L$ should again scale as $\Omega(N/L^2)$. To keep the NL almost
constant (independent of $N$), one should be prepared to scale the
average power and bandwidth at least linearly with the network
size.

The above bounds lead to the question of designing efficient
decentralized routing and MAC protocols. We propose
Multi-Resolution Randomized Hierarchy (MRRH), a novel randomized
algorithm for efficient wireless ad-hoc networking and measure
various performance metrics of MRRH.  Nodes in an MRRH network
manage their communication strategies locally. Even more
interestingly, {\em network evolution in MRRH is totally
stateless}, that is, nodes randomly and independently change their
strategies regardless of all other nodes in the network, making
MRRH a perfect candidate for mobile environments. The proposed
{\em routing algorithm defined on the underlying random network is
also stateless}: A node will decide on where to relay a received
packet based only on its own position, the position of the target,
and the positions of at most $O(\log N)$ neighboring nodes. Even
though the routing is stateless, the NL of MRRH, in terms of the
average number of hops per communication, is only $O(\log^3 N)$.
Furthermore, the power and bandwidth requirement of MRRH are
nearly optimal among all routing algorithms that have an average
NL equal to the one provided by MRRH, as is explained next.

The main idea behind MRRH is to superimpose several virtual
topologies to form a nested hierarchical structure. At all times,
nodes that belong to a higher hierarchy are also members of all
lower hierarchies. The average distance of communication in upper
hierarchies is exponentially longer than the ones in lower
hierarchies. The fraction of times a node is part of a higher
hierarchy, however, is exponentially smaller. A packet is usually
relayed starting from a low hierarchy. If the target node is far away, then
the packet automatically \emph{climbs up} the hierarchy and
quickly reaches a node in upper hierarchies. These hierarchical
structures are not constant, and will change from time to time or
even from packet to packet. Nodes that operate in higher
hierarchies for some routes might be part of lower hierarchies for
others. This will provide a natural load balance to the system.
The overall algorithm however, guarantees  correct delivery of all
packets even in this highly changing environment.

\subsection{Relation to Previous Work}
Our routing algorithm is based on position. Various position-based
routing algorithms have been proposed for WANET's. For the purpose
of this paper, these algorithms can be divided into two main
categories \cite{surv}:\\
\emph{(i) Nearest neighbor, approximate line of sight, routing:}\\
Algorithms based on nearest neighbor communication pass on a data
packet to a close by node which is closer to the destination.
Various variations on this theme can be found in
\cite{NK,HL,TK,LS,SL,SL2} and many other papers. Nearest neighbor
communications incurs a large NL, often inappropriate for large
scale operations. In a network of size $N$ with nodes randomly
distributed on a square, nearest neighbor communication requires
an $\Theta (\sqrt{N})$ hops for an average
communication. To reduce the NL, various hierarchical routing algorithms have been considered. \\
\emph{(ii) Hierarchical routing algorithms:}\\ In these algorithms,
the routing is done in (usually two) different levels. For instance,
zone-based routing algorithms divide the network to various zones.
The routing is divided into two steps, routing between the zones,
and routing within the zones by introducing  a set of ``dominating
nodes''. A dominating node is able to reach any node within a zone
and also is capable of communicating with dominating nodes in other
zones. Variations of this idea have been proposed in many papers
including \cite{JL,LAR,WL,BMSU,DSW}. Since there are fewer
dominating nodes compared to all nodes, the number of hops for an
average communication reduces. While these two level hierarchical
schemes mitigate the NL problem to some extent in medium size
applications, they will not scale appropriately to extremely large
network sizes.

MRRH too is a hierarchical system, except that the number of
hierarchal levels is not constant, and that nodes frequently join
and leave hierarchies through local decisions. The overall design
of MRRH however, ensures that, (i), the network is totally
connected within each level of hierarchy at all times, and (ii)
there is always a polylogarithmically \footnote {A function $f(N)$
is said to be polylogarithmic, if there exist $m,M>0$ such that
$f(N)<\log^ m(N)$ for all $N>M$. } small path between any pair of
nodes, that might pass through various hierarchical levels, and
(iii) this path can be discovered locally through a simple greedy
algorithm. A greedy routing algorithm will then be able to
efficiently route data packets from any source to any destination
within small latencies.

The rest of this paper is organized as follows. In Section
\ref{sec:pre}, we introduce the problem model.  In Section
\ref{sec:bounds}, a set of new general constraints are derived
that relate the maximum feasible throughput to the average power,
bandwidth and NL.  MRRH along with our proposed routing algorithm
is introduced in Section \ref{sec:mrrh}. Then in Section
\ref{sec:power}, power and bandwidth requirement of MRRH are
derived. Comparison with results in Section \ref{sec:bounds}
enables us to prove that MRRH is nearly optimal in its bandwidth
and power usage. Section \ref{sec:conc} provides concluding
remarks.

\section{Preliminaries}\label{sec:pre}

This section formally introduces the WANET model used in this paper,
including the network topology, traffic and communication models.

\subsection{Network Model} The wireless network model considered
in this paper consists of a set of $N$ nodes, randomly distributed
on the surface of a sphere of radius $R$ and area $A=4\pi R^2$. We
denote the set of all nodes by $\Gamma$. Symmetric surface of the
sphere simplifies geometric proofs of the paper. Most of the proofs
however can apply to a regular two dimensional geometry (e.g., a
square) with only simple modifications.

Each node can act both as a transmitter and a receiver. Every node
has access to $K$ non-interfering Additive White Gaussian Noise
(AWGN) channels. The bandwidth of channel $k\in {\cal K}\triangleq
\{0,1,2,...,K-1\}$ is assumed to be $B^{k}$ and the noise power
spectral density is $\eta_0$ for all channels. 
  At each channel and
at any point of time, a node is assumed to be transmitting data to
or receiving data from at most one other node. Therefore, time
sharing has to be used for communication to multiple nodes over a
single channel. For simplicity, it is assumed that the communication
at each channel is performed with a common power $P^k$, i.e., if a
node decides to transmit on channel $k$, it will do so with a power
$P^k$. The participation function, $\phi^k_t(i,j)$, is one if node
$i$ is transmitting to node $j$ over its $k^{th}$ channel at time
$t$ and is zero otherwise. With this definition, at a given time
$t$, a node $i$ is assumed to be able to communicate to a node $j$
with rate $R^k_t(i,j)$ equal to the capacity of the corresponding
AWGN channel:
\begin{eqnarray}\label{eqn:uwbr}
R^k_t(i,j)=B_k\log_2\left(1+\frac{P^{k}\gamma(||X_i-X_j||)}{B_k
\eta_0 +\sum_{l\neq j}
P^{k}\phi^k_t(j,l)\gamma(||X_j-X_l||)}\right)\nonumber
\end{eqnarray}

where $X_x$ is the position of a node $x$ and $||.||$ is the
geodesic distance on the surface of the sphere and
$\gamma:{\mathbb R}^+\rightarrow (0,1]$ is a power dissipation
function.

Average power consumption at a node $i$ is:
\begin{equation}\label{eqn:ap}
P_{avg}(i)=E_t\left\{\sum_{j\in \Gamma} \phi^k_t(i,j)
P^k_t(i,j)\right\}
\end{equation}

where $E_t\{.\}$ denotes time averaging. $P_{avg}=N^{-1} \sum_{i\in
\Gamma} P_{avg}(i)$ is the total average power. The total bandwidth
of the system is $B=\sum_{k=0}^{K-1} B^k$.

For a fixed $P_{avg}$ and $B$, the choice of the power and bandwidth
levels $P^k, B^k$ are left to the WANET designer.

\subsection{Multi-Hop Routing} Data routing is performed through
a decentralized multi-hop algorithm. Upon receiving the data packets
for a destination node, a relay node should be able to locally
decide on where to send the packet next. Ideally, the routing
algorithm should be stateless, that is, (1) the data packet should
only contain information about the destination and possibly the
source node and (2) routing decisions should be made on a ``per
packet'' basis. In our model we assume nodes are equipped with
Global Position Systems (GPS) and addressing is by position, i.e.,
packets contain the position of the destination node.  A routing
algorithm is said to have an average NL of $L$ if each packet, on
average, has to be relayed $L$ times before it reaches the
destination, where the averaging is done over nodes and time.

\subsection{Traffic Model} We assume a uniform and symmetric
traffic model. At any given time, any node $i$ is sending packets
to exactly one node $j$ at a rate of $\lambda$ bits per second,
called the throughput. We call a throughput $\lambda$ feasible if
it can be relayed successfully at all nodes using a finite buffer
size.

 Equivalently, a throughput $\lambda$ is feasible with a
participation policy function $\phi^k_t(i,j)$ if and only if:
\begin{eqnarray*}
\forall i\in \Gamma,\;\;\;\sum_{k\in {\cal K}}\left(
E_t\{\sum_{j\in \Gamma}\phi^k_t(i,j) R^k_t(i,j)\}- E_t\{\sum_{l\in
\Gamma}\phi^k_t(l,i) R^k_t(l,i)\}\right) \geq 0
\end{eqnarray*}

where $E_t$ denotes expectation over time, and we have considered
the fact that every node is a sink of data with rate $\lambda$ and
the source of some data with the same rate $\lambda$.

In this paper, we will be interested in minimizing the average NL
$L$ while maximizing  a feasible throughput $\lambda$ and
minimizing the average power consumption $P_{avg}$ and bandwidth
requirement $B$. These requirements are of course conflicting. A
set of bounds are derived in Section \ref{sec:bounds} that
quantifies these conflicts. We then propose a system capable of
efficiently trading off these conflicting figures of merits. The
proposed system, called Multi-Resolution Randomized Hierarchy
(MRRH) is introduced in Section (\ref{sec:mrrh}) and is shown to
be nearly optimal in the light of the bounds derived in Section
\ref{sec:bounds}.

\section{Power and Bandwidth Requirement of Low NL WANETs}\label{sec:bounds}
In this section, we derive two new lower bounds on the average
power and total bandwidth requirement of \emph{any} WANET with a
feasible constant throughput of $\lambda$ and average NL of $L$.

\subsection{A Lower Bound on Average Power}
Reducing the average NL requires and increase in the average
communication length which imposes a constraint on the average
power required for communication. We quantify this requirement in
this subsection.

 We first prove the following
simple Lemma.
\begin{lem}\label{lem:apart} On average, $N/16$ of all communication pairs are between nodes
that are at least a distance $R/4$ apart. Therefore, at least an
average of $\lambda/16$ bits per second should be communicated
between nodes that are at least a distance $R/4$ apart.
\end{lem}

\begin{proof}

This is easily proved as follows. Consider two caps on the poles
of sphere with angles $\theta_1 \in (\pi/4, \pi/2)$ and $\theta_2
\in (-\pi/2, -\pi/4)$. Each of these caps has an area $R^2
\sqrt{2}/2 \geq A/4$ and therefore contains on average $N/4$
nodes. The probability that two nodes, one each cap, communicate
in any given round is therefore at least $(1/4)^2=1/16$. The
minimum distance between
these two nodes is $R/4$.\\
\end{proof}

 Let
$\Gamma^T_t, \Gamma^R_t$ indicate the set of nodes that are
transmitting to and receiving data from other nodes that are at
least $R/4$ away. We have shown that
$E_t\{|\Gamma^T_t|\}=E_t\{|\Gamma^R_t|\}\geq N/16$. Note that when
the communication latency is at most $L$, each packet starting
from a node in $\Gamma^T_t$ and ending in a node in $\Gamma^R_t$
must pass over a link that is at least $R/(4L)$ long.

Lemma \ref{lem:apart} leads to the following Theorem on average
power consumption of any MANET with average NL $L$.

\begin{thm}\label{prop:power} Any routing algorithm with feasible throughput
$\lambda$ and average NL $L$  requires an average power of at
least $P_{avg}> \frac{\eta_0  \lambda }{16 \gamma(\frac{R}{4L})\ln
2}$. For a physical power dissipation ($\gamma(D)\propto D^{-2}$),
and the node density a constant $\rho$, one has $P_{avg}\geq
\lambda \rho^{-1} \eta_0^{-1}48^{-1} (4\pi)^{-1} N L^{-2}$.
\end{thm}

\begin{proof}
It is easy to show that:
\begin{equation}\label{eqn:pwr}
R^k_t(i,j) \leq \ln 2 \, P^k \gamma(||X_i-X_j||)\phi^k_t(i,j)
/\eta_0
\end{equation}
by letting $B^k  \rightarrow \infty$ in Eqn. (\ref{eqn:uwbr}).
Consider a data stream of rate $r_{xy}$ bits per second, starting
from a node $x\in \Gamma^T_t$ that reaches a node $y \in
\Gamma^R_t$ using only $L$ steps. Data packets have to be
communicated over a link that is at least $R/(4L)$ long. This
requires a power of at least $P_{xy} \geq r_{xy} \eta_0\ln
2/\gamma(R/(4L))$. The total power required for accommodating all
communications between nodes in $\Gamma^T_t,\Gamma^R_t$ only is at
least:
\begin{eqnarray*}
P_{avg}N &\geq& \sum_{x \in \Gamma^T_t, y\in \Gamma^R_t}
P_{xy}\\&\geq&  \frac{ \eta_0}{\gamma(R/(4L))\ln 2} \sum_{x \in
\Gamma^T_t, y\in
\Gamma^R_t} r_{xy}\\
&\geq &\frac{\eta_0 N \lambda }{16 \gamma(\frac{R}{4L})\ln 2}
\end{eqnarray*}

which proves the claim.
\end{proof}

\subsection{Lower Bound on Total Bandwidth}

Theorem (\ref{prop:power}) proved that for a physical system with
almost constant NL $L$, the average power should scale at least
linearly with $N$.  We next show that a system with low NL
requires a considerable amount of bandwidth to scale.

The idea is that , from Lemma \ref{lem:apart}, to provide a low
NL, an average communication has to use long range communication
on one of the various channels. A long range communication in a
channel will significantly interfere with most of the nodes in the
same channel , disabling them from simultaneous communication.
This in turn limits the amount of data that can be mobilized in
the network. This is quantizes in the following Theorem.

\begin{thm}\label{prop:bnd} For any system with average NL of $L$ and a
feasible throughput of $\lambda$ one requires that:
\[
B\geq \frac{\gamma(2\pi R)}{\gamma(\frac{R}{4L})}\times
\frac{N\lambda}{16 K \eta_0}
\]\\
\end{thm}

\begin{proof}

Consider  $B_k$ the bandwidth of some channel $k $. At the time $t
$, let's define $P _{R}(i)$ as the signal power received by node
$i$ over channel $k $ \emph{corresponding to a long range
communication only}. If $i$ dose not communicate in $k $ at $t $,
that is, if $i\notin \Gamma^k_t$, then $P _{R}(i)=0$. Similarly,
let $P _{I}(i)$ represent the interference of all other \emph{long
range communications} at node $i$ at time $t $.

Let ${\cal R}_k (i)$ denote the average communication rate a node
$i$ receives through long range communication. Then:

\begin{eqnarray}\label{eqn:upper}
{\cal R}_k (i)&\leq & B _k \log \left(1+\frac{P_R(i)}{B
_k\eta_0+P_I(i)}\right)\nonumber\\
&\leq& \frac{B _k P _{R}(i)}{\eta_0B _k+P _{I}(i)}\nonumber\\
&\leq& \frac{B _k P _{R}(i)}{\eta_0B _k+\sum_{j\neq
i}\frac{\gamma(2\pi R)}{\gamma(R/(4L))}P _{R}(j)}\nonumber\\
&=&B _k c^{-1} \frac{c P _{R}(i)}{\eta_0B _k+c \sum_{j\neq i}P
_{R}(j)}
 \end{eqnarray}

where we have defined $c\triangleq \frac{\gamma(2\pi
R)}{\gamma(R/(4L))}$. The last inequality is understood as
follows: suppose node $j$ receives a power $P$ corresponding to a
long range communication from a node $j$. By definition, node $j$
is at least $R/(4L)$ apart from $i$. Therefore, the power is has
used for communication must be at least $P/\gamma(\frac{R}{4L})$.
This power imposes at least $\gamma(4\pi R)P/\gamma(\frac{R}{4L})$
interference on all nodes other than $i$.\\

The total communication rate over long range connections is
bounded as:
\begin{equation}\label{eqn:mmm}
\sum_{i} {\cal R}_k (i) \leq B _k c^{-1}\sum_i  \frac{c P
_{R}(i)}{\eta_0B _k+c \sum_{j\neq i}P _{R}(j)}
\end{equation}

The right hand side of the above inequality can be maximized by
noting that:

\begin{eqnarray*}
H&\triangleq&\sum_i  \frac{c P _{R}(i)}{\eta_0B _k+c \sum_{j\neq
i}P _{R}(j)}\nonumber \\
&=&\sum_i \left(\frac{c P _{R}(i)+\eta_0B _k+c \sum_{j\neq i}P
_{R}(j)}{\eta_0B _k+c \sum_{j\neq
i}P _{R}(j)}-1\right)\nonumber\\
&=& \sum_i \frac{\eta_0 B _k +c \sum_j P _{R}(j)}{\eta_0B _k+c
\sum_{j\neq i}P _{R}(j)}- N\nonumber \\
&=& \sum_i \frac{\eta_0 B _k +c T _R}{\eta_0B _k+c \sum_{j\neq i}P
_{R}(j)}-N \nonumber
\end{eqnarray*}
where $T _R$ is the total received signal power used for long
range communication.

Now let's make the following change of variable: $Q
(i)=\sum_{j\neq i} P _{R}(j)$. Then, :
\begin{eqnarray}\label{eqn:min}
H&=&\sum_i \frac{\eta_0 B _k +c T _R}{\eta_0B _k+c Q (i)}-N
\end{eqnarray}

Now note that $\sum_{i} Q (i)= (N-1) \sum_i P _R (i)= (N-1) T _R$.
For a fixed $T _R$, maximizing (\ref{eqn:min}) under this
constraint will require $Q (i)= (N-1)T _R /N$, which results in:

\begin{eqnarray}\label{eqn:ming}
H&\leq &N \frac{\eta_0 B _k +c T _R}{\eta_0B _k+c
\frac{(N-1)}{N}T _R}-N\nonumber\\
&=&N \left(\frac{\eta_0 B _k +c T _R}{\eta_0B _k+c
\frac{(N-1)}{N}T _R}-1\right)\nonumber\\
&=& N \left(\frac{c \frac{T _R}{N}}{\eta_0B _k+c
\frac{(N-1)}{N}T _R}\right)\nonumber\\
&\approx& c\frac{ T _R}{\eta_0B _k+c T _R}
\end{eqnarray}

Inserting this back into (\ref{eqn:mmm}):
\[
\sum_{i} {\cal R}_k (i)\leq \frac{ B _k T _R}{\eta_0B _k+c T _R}
\]

The above inequality holds for any channel $k $. We know that the
rate of communication over long range connections should be at
least $N\lambda /16$ (Lemma \ref{lem:apart}).  Therefore, we need
to have $\sum_{k \in {\cal K}}\sum_{i} {\cal R}_k (i)\geq
(N\lambda)/(16\eta_0\ln 2)$. Therefore, there should exist at
least one channel $k^*$ such that,
\[
\sum_{i} {\cal R}_{k^*} (i)\geq (N\lambda)/(16\eta_0\ln 2)/K
\]

For this channel,

\[
\frac{ c B _{k^*} T^* _R}{\eta_0B _{k^*}+c T^* _R}\geq c
(N\lambda)/(16\eta_0)
\]

Noting that all components are positive, the necessary condition for
the above inequality to hold is that:
\[
T^* _R \geq (N\lambda)/(16K \eta_0 \ln 2)
\] and that
\[
B _{k^*} \geq c (N\lambda)/(16K \eta_0)= \frac{\gamma(2 \pi
R)}{\gamma(\frac{R}{4L})}\frac{N\lambda}{16 K \eta_0 \ln 2}
\]

\end{proof}


The next section introduces our low NL MRRH solution. We provide
an algorithmic description of MRRH, prove its correctness and NL
properties and calculate its power and bandwidth requirements.
Equipped with the results in this section, we will be able to show
that MRRH requires an almost optimal power and bandwidth among all
systems that provide the same average NL as MRRH does.

\section{Multi-Resolution Randomized Hierarchy}\label{sec:mrrh}

Multi-Resolution Randomized Hierarchy (MRRH) is a joint
participation policy and routing algorithm for efficient delivery
of data packets, as follows.

MRRH uses $K=\log N- 2\log\log N $ different channels, where logs
are all in base $2$ unless otherwise specified. Let $\angle (i,j)$
denote the spherical angle between two nodes $i,j$. Take a cap on
the sphere with spherical angle $\theta_k$ and surface area
$A_k=2\pi R^2 (1-\cos \theta_k)$ such that $A/A_k=(16 \log N)^{-1}
2^{-k} N $, where $A=4 \pi R^2$ is the total surface of the
sphere. Caps of angle $\theta_k$ will determine the communication
neighborhood of MRRH at channel $k$. In other words, two nodes
$i,j\in\Gamma$ will communicate in channel $k$ only if
$\angle(i,j)<\theta_k$ in which case we say $i,j$ are neighbors in
the $k^{th}$ channel.

 We call $A_k$ the coverage area of channel
$k$. The coverage area as well as the communication range grow
exponentially with the channel level $k$. For simplicity, we
define the set of neighbors of $i$ in the $k^{th}$ channel as
$\Psi^k(i)$.

 Note that, $A_k=2\pi R^2 (1-\cos \theta_k) \leq 2\pi
R^2 \theta_k^2$.  Thus, for any two nodes $i,j$ that are
\emph{not} neighbors in channel $k$ we have:

\begin{eqnarray}\label{eqn:rk}
||X_i-X_j||&\geq&  2\pi R \theta_k \geq 2\pi \sqrt{\frac{A_k}{2\pi
R^2}}\\&=&\sqrt{2\pi A_k}\nonumber\\&=& \sqrt{2\pi\cdot 4\pi
(16\log N)2^k N^{-1}}\nonumber \\&=&8\pi R \sqrt {2^k N^{-1} \log
N}\nonumber
\end{eqnarray}

We will use the above fact to upper bound the amount of
interference a node's communication incurs on other nodes that are
not in its neighborhood.

The participation policy of MRRH, $\phi^k_t(i,j)$, is randomized.
Each node $i$ will randomly choose a value $0\leq l(i)\leq K$,
called its level, as follows: $Pr\{l(i)=0\}=1/2$ and
$Pr\{l(i)=k+1\}=Pr\{l(i)=k\}/2$ for $0<k<K-1$. For $k=K$, the
probability is $Pr\{l(i)=K\}=1/2+(1/2)^{K-1}$. Each node $i$ with
level $l(i)$ will only open its first $l(i)$ channels, i.e., it will
only communicate on channels $k=0,1,...,l(i)$. Note that if channel
$k$ of any node $i$ is open, all lower channels ($0,1,...,k-1$) of
$i$ are also open.

Evidently, the above participation policy is a homogenous one, i.e.,
it treats all the nodes uniformly, but imposes a heterogenous
structure by placing nodes randomly at different levels (or
hierarchies). The density of the nodes decreases exponentially as
their level increases. As will be seen shortly, nodes in upper
levels will have to spend an exponentially larger amount of power.

When the nodes in the WANET have heterogenous power capabilities,
the level of nodes can be aligned with those capabilities. In other
words, nodes can choose a level between $0$ and $K$ based on their
power capabilities. All the results in this section remain unchanged
provided that the density of nodes in different levels follows the
same exponential relation. For clarity, however, we state all
results for a homogenous setting.

The routing is greedy. When a node $i$ receives a packet destined
for a node $j$, it checks the neighborhood of all its \emph{open}
channels and sends the packet to a neighbor that is closest to the
destination node $j$ and is also closer to $j$ than $i$ itself. If
no such node exists, the routing is stopped. The routing  either
stops at the target node  $j$ (successful routing) or at a wrong
node $j'\neq j$ (unsuccessful routing).

The reader familiar with data structures and algorithms has
probably noticed the similarity of MRRH with SkipLists
\cite{skiplist}. SkipLists are randomized versions of binary
search trees. MRRH has many similarities to SkipLists in its
hierarchical structure and greedy search strategy. SkipLists
however do not have any notion of geographical location,
dimension,  power, bandwidth and interference.

This section will prove various routability properties of MRRH.
These routing properties are geometric in nature and do not take
into account physical requirements of  routing (i.e., rate, power
and bandwidth requirements) . Rather, they are only concerned with
proving the correctness of the data delivery and calculating the
average and maximum NL. Following sections, on the other hand,
will find the power and bandwidth required by MRRH for making a
desired throughput feasible. There, for instance, we will employ a
simple Time Division Multiple Access (TDMA) scheme for shared
channel access by neighboring nodes.

Throughout, an event is said to happen with high probability
(w.h.p) if it happens with probability at least $1-N^{-2}$. The
correctness of routing for MRRH can be proved by methods close to
continuum percolation arguments \cite{cont-perc,gupta-perc}. We
adopt a different approach for proving this. Taking any pair of
nodes $i,j$, we will show that there exists a chain of nodes
$s_1=i\rightarrow s_2 \rightarrow ... \rightarrow s_m =j$ such
that $s_{n+1}\in \Psi^k(s_n)$ and $\angle (s_{n+1},j)<\angle
(s_n,j)-\epsilon_k$, for some constant $\epsilon_k>0$. In other
words, there is a chain of neighbors that will take any packet
starting from $i$ to $j$ and the packet gets strictly closer to
$j$ at each step (Lemma \ref{lem:l1}).

For that we need a number of simple lemmas.

\begin{lem}\label{lem:chern} Any surface of area $S= 2^{k-2} A_0$ w.h.p. contains at
least $2\log N$ nodes and at most $6 \log N$.
\end{lem}
\begin{proof} This follows easily from Chernoff bound. Note that the average
number of nodes in $S$ is $\mu=N S/A=4\log N$. By Chernoff bound,
the probability that the actual number of nodes is not in the
interval $(0.5\mu,1.5\mu)=(2 \log N, 6 \log N)$ is at most
$e^{-(1/2)^3\times 16\log N}<N^{-2}$.
\end{proof}

The following is a simple  trigonometric inequality, the proof of
which is  omitted.
\begin{lem}\label{lem:ref}
$\forall \theta\in (0,\pi]$,
$1/4<\frac{1-\cos(\theta/2)}{1-\cos(\theta)}<1/2$.
\end{lem}

\begin{figure}
\begin{center}
\includegraphics[height=2.6in, width=2.1in]{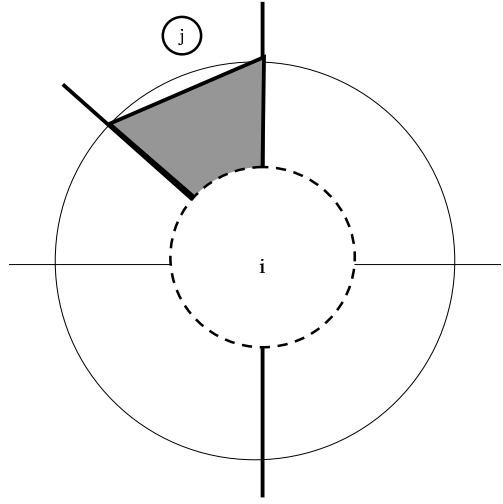}
\caption{Current relay node $i$ and its relative position to
destination node at  $j$. Angular radius of the larger circle is
$\theta_k$ where as the radius of the dotted circle is $\theta_k/2
$.}\label{f1}
\end{center}
\end{figure}

Next,

\begin{lem}\label{lem:l1} Assume that a packet destined for a node $m$
is currently at a node $i\neq m$. For any open channel $k\leq l(i)$
of $i$, either $m\in \Psi^k(i)$ in which case the packet will be
directly forwarded to the target $m$ through channel $k$ or else
 w.h.p. there exists a node $s\in \Psi_k$ that is
$\epsilon_k/2$ closer to $m$ than $i$, that is: $\angle (s,m)\leq
\angle (i,m)-\epsilon_k/2$.
\end{lem}
\begin{proof}  If $m\in \Psi^k(i)$ the the packet will of course be sent to $m$ directly. Otherwise, consider two caps of angles $\theta_k, \theta_k/2$  centered at the current relay node $i$. Without loss of generality, assume the target is in the first
octant around the node $i$ as in Fig. \ref{f1}. Therefore any node
in area indicated by $B$ is closer to $m$ than the node $i$. The
area of $B$ is  $2\pi R(1-\cos(\theta_k/2))\geq (1/4)2\pi
R(1-\cos(\theta_k))=A_k/4=2^{k-2}A_0$,  by using Lemma
\ref{lem:ref}. By Lemma \ref{lem:chern} on the other hand, this
area contains at least $2\log N$ nodes w.h.p., that have their
$k^{th}$ channel open. Each of these nodes is closer to $m$ than
$i$. Therefore, the packet will get an angle of at least
$\epsilon_{k}=\theta_k/2$ closer to the target w.h.p.
\end{proof}

Lemma \ref{lem:l1} can now be used to prove the correctness of the
routing in MRRH.

\begin{thm}\label{pr:1}
The probability that there exists any unroutable pair of
communication in the network is at most $N^{-1}$.
\end{thm}
\begin{proof} The proof follows from the fact that the base
channel of all nodes are always open. Applying Lemma \ref{lem:l1}
to the base channel shows that the message gets an angle of least
$\theta_0/2\geq (64\pi \log N)/N$ degrees closer to the target at
each hop, and thus it takes at most $\frac{N}{32 \pi \log N}$ hops
for any packet to be delivered. The probability of the failure is
by union bound at most $N^{-2}\times N/(32 \pi \log N)<N^{-1}$.
\end{proof}

\smallskip

  The correctness of MRRH is not enough for that to
perform efficient data delivery. Average NL of MRRH is in fact
only $O(\log^3 N)$, as is discussed next. The idea behind this
result is the following: consider a packet starting from $i$ and
heading for a node $j$. If $j$ is far from $i$, the packet will be
communicated by means of links in upper channels (larger $k$)
using longer range hops. As such, the packet gets quickly close to
$j$. Once it gets to the vicinity of $j$, the routing will use
shorter range connections.

 The participation function of the MRRH,
and the connectivity of MRRH topology at each level, ensure the
success of the algorithm. The key ingredients of the proof are
that if a packet is further from destination than the range of a
channel $k$, and if the current node has a channel $k'>k$ open,
the packet will always be routed through channel $k'$ or a higher
channel. Therefore, the packet  \emph{climbs up} the hierarchy,
exploiting long range connections provided by nodes in the upper
hierarchies to get  close to its destination quickly. The
structure of the MRRH ensures that finding nodes in upper
hierarchies is always possible for all packets starting from any
node in the network.

Let $C(i,j)$ denote the cap centered at $i$ with angle $\angle
(i,j)$. In other words, $C(i,j)$ is the cap centered at $i$ that
passes through $j$. Let $A(i,j)=A(j,i)=2\pi R^2 (1-\cos (\angle
(i,j)))$ denote the area of this cap.

We now move to the main Theorem of this section.
\smallskip

\begin{figure}
\begin{center}
\includegraphics[height=2.0in, width=2.5in]{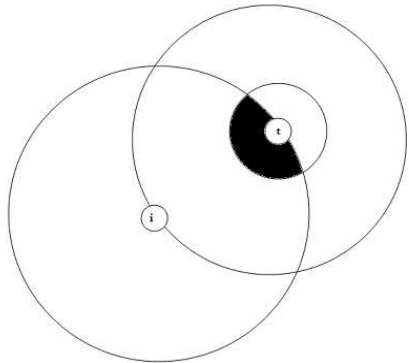}
\caption{The ideas in Theorem \ref{p2}: When node $i$ is in phase
$g$ of the target, the solid area is the one reached by all nodes
in phase $g-1$ of the target.}\label{f2}
\end{center}
\end{figure}

\begin{thm}\label{p2} The average NL of  MRRH is $L_{avg}< 2 \log^3 N$.
\end{thm}
\smallskip
\begin{proof} Consider a packet heading towards a target node $j$. We say that routing is
in phase $g$ if the current position of the packet is node $i$ and
$\theta_g<\angle (j,i) \leq \theta_{g+1}$, or equivalently, when
$A_g < A(i,j)\leq A_{g+1}$. The starting phase is at most $\log
N$. When the packet reaches phase $0$, it can be delivered to the
target immediately though the base channel. We now show that after
$2\log N$ routing steps, the phase of the packet is decreased by
at least $1$ w.h.p.

To see this, assume a communication to be in phase $g>1$. First
assume that $l(i)\geq g+1$, that is, $i$ has its $g+1^{th}$
channel open. Therefore, it can reach an area of size at least
$A_k/4=2^{g-2}A_0$ of all nodes in phase $g-1$ of the target (see
Fig. \ref{f2}).  From lemma (\ref{lem:chern}), this area contains
at least $2\log N$ nodes whose channel $g+1$ are open w.h.p.
Therefore, if $l(i)>g+1$, w.h.p. the message will be passed to a
node in phase $g-1$ of the target.

If $l(i)<g+1$, then by the routing algorithm, it passes the packet
to the closest node to the target. From (\ref{lem:l1}) we know
that there is at least one node closer to the target than $i$
w.h.p., hence the phase of the routing is never decreased.

 There remains to show that for any $g>0$, after at most $2 \log(N)$
steps, any packet can arrive at a node operating at a level
greater than or equal to $g+1$. The worst case is when the node in
the starting position operates at channel $0$ only, and the target
is furthest ($\theta=2\pi$ ) apart. It can be easily verified that
by at most $2\log(N)$ sampling of nodes working in channel $g$ a
node whose channel $g+1$ is open can be found w.h.p., where as
this takes two steps on average. If the packet is already in phase
$g+1$ we are finished. Since going to an upper channel takes less
than $2\log(N)$ steps and there are only $K$ such channels, the
total number of steps required to find a node in any channel is at
most $2K \log N$ steps w.h.p. With these many steps, the phase of
the routing is decreased by at least $1$ w.h.p. The maximum phase
to consider is
 $K<\log N $. Thus, any packet will reach the
destination in at most $2K \log^2 N$ with high probability. Noting
that $K<\log N$, we get the desired result.
\end{proof}

Along the lines of proof of Theorem \ref{p2}, we have also proved
the following lemma that will prove useful in bounding the
bandwidth
requirement of MRRH nodes in later sections,\\
\begin{lem}\label{lem:co1} If a message is being delivered to a target
from channel $k$ of a node in phase $g>k$ of the target, it almost
surely leaves channel $k$ for an upper channel after at most
$2\log(N)$ consecutive moves in channel $k$, independent of $k$.
\end{lem}

So far, we have proved various routing properties of MRRH. In
particular, we have shown that MRRH is routable, in the sense that
any given packet from any node to any destination will be delivered
in finite time w.h.p. More over, we have shown that the number of
hops taken for the delivery of each packet is on average at most
$2\log^3 N$.

These observations are all geometric in nature. In the following
sections, we derive the feasible throughput of MRRH for a given
average power and total bandwidth.

\section{Power and Bandwidth Requirement of MRRH}\label{sec:power}

In this section, we derive the power and bandwidth required by
MRRH. To do so, we need to upper bound the rate of the
communication required at any given channel. We then calculate the
power and the amount of bandwidth required for communication in
each channel. Finally, by comparison with results in propositions
\ref{prop:power}, ~\ref{prop:bnd}, we show that the power and
bandwidth requirements of MRRH are only polylogarithmically away
from the lower bounds.

\subsection{Bit Rate Requirement at Different Channels}
We start by finding the rate with which each node $i$ has to
communicate with its neighbor nodes $\Psi^k(i)$ at each channel $k$
in order to make a total throughput $\lambda$ feasible. To do this,
we need to upper bound the probability that channel $k$ of a random
node $i$ is used for a random routing, assuming that $l(i)\geq k$
(i.e., channel $k$ of $i$ is open). This will determine the amount
of load that has to be handled by channel $k$ of node $i$.

The next lemma shows that a packet does not \emph{stay} in any
channel $k$ for a long time.

\begin{lem}\label{l31} Assume a packet, targeted for a node $j$, is routed to a node $i$ through its
$k^{th}$ channel. If $j \notin \Psi^k(i)$, the next routing will
happen in a  channel $k'\geq k$ w.h.p.
\end{lem}

\begin{proof} This lemma states that the channel in which the packet is routed will decrease only when the packet gets close
enough to the final target. The idea of the proof is that if the
target is far, w.h.p. a node closer to it in channel $k$ will be
found and thus the level of the node receiving the packet does not
decrease. The details of the proof are omitted.
\end{proof}

\begin{lem}\label{l3}
Let $U^{k}$ be the number of times the $k^{th}$ channel of a
random node $i$ is used for $N/2$ random packets. Then
$U^k<2^{k+2} \log N$ w.h.p.
\end{lem}
\begin{proof}
 Suppose a packet has started from a random node $c$ towards a random destination $j$ and is currently being relayed by a node $i$ over its $k^{th}$ channel.
  Two scenarios are possible: (1) If $i$ is in phase $g\geq k+1$ of the target, then through Lemma \ref{l31}
 the message will leave channel $k$ after at most $2\log N$ more steps  in channel $k$ w.h.p.. (2) If the message is in a phase $g< k+1$, then
 through Lemma \ref{l31}, it will immediately leave channel $k$ w.h.p.
Thus, the necessary condition for a channel $k$ of any node $i$ to
be involved in the transmission of a
 packet is that either the target is in its neighborhood, i.e.,  $j\in \Psi^k(i)$, or by lemma \ref{lem:co1}, $\angle (i,c)< 2\log(N)
 \theta_k$.

 Let $p_k$ be the probability that the $k^{th}$ channel of a
 randomly chosen node $i$ participates in a random communication
 from a
 randomly chosen source $s$ to a destination $t$. By the union bound, $p_k$ can  be
 bounded as:
 \begin{eqnarray*}\label{b1}
    p_{k}&\leq &Pr\{\angle (i,s) <2\theta_k \log N  \}
    + Pr\{t\in \Psi^k(i)\}
    \\
    &\leq &4 \times 2^k \log N /N
\end{eqnarray*}
For $N/2$ communication pairs, the average number of times
communication pairs for which channel $k$ of node $i$ has to relay
packets is $2^{k+1} \log N$. Applying the Chernoff bound gives the
result.
\end{proof}

We can now use Lemma (\ref{l3}) to upper bound the average
communication rate required for making a uniform throughput of
$\lambda$ feasible.\\

\begin{thm}[\textbf{Bit-Rate}] \label{prop:wk}
In MRRH, a communication rate of $R^k(i,j)=24 \lambda \log^2 N 2^k$
for any $k\leq K$ is sufficient for a uniform throughput of
$\lambda$ to be feasible. This can be achieved using a Time Division
Multiple Access strategy.
\end{thm}
\begin{proof} From Lemma \ref{l3}, for $N/2$ random source-destination communications, the maximum number of data packets
that has to be communicated through a channel $k \leq K$ of any node
$i$ whose level $l(i)$ is greater than $k$ is $2^{k+2} \log N$. For
a throughput of  $\lambda$ bits/sec, this means that data with the
rate of at most $\lambda 2^{k+2} \log N$ bits per second has to be
communicated through channel $k$.

To resolve the conflicts we adopt a simple TDMA approach. We assume
that when a node $i$ needs to communicate in channel $k$, no other
node in its neighborhood, $\Psi^k(i)$, will communicate in channel
$k$. But from lemma \ref{lem:chern}, there are at most $6 \log N$
such nodes. This TDMA scheme can thus be used to yield communication
turn to the nodes by slotting time into at most $6 \log_2 N$ slots
in each channel. To communicate $\lambda 2^{k+2} \log N$ bits per
second, the communication rate thus needs to be at most: $R^k(i,j)=
24 \lambda 2^{k} \log N $ for any two nodes $i,j$ communicating in
channel $k$.\\
\end{proof}

\subsection{Power Requirement}

Using Eqn. (\ref{eqn:uwbr}) we can find the power $P^k$ necessary
 for providing the bit-rate required in
(\ref{prop:wk}). For now assume that the bandwidth at each channel
is infinite, i.e., $B_k\rightarrow \infty$ for all $k\in {\cal
K}$. Eqn. (\ref{eqn:uwbr}) will therefore reduce to a linear
relation between the transmitted power $P^k$ and the bit-rate
$R^k$:
\begin{equation}\label{eqn:binf}
R^k(i,j)=\ln 2 P^k \gamma(||X_i-X_j||)/\eta_0
\end{equation}

Let's assume a specific power decay function of the form\\
$\gamma(||X_i-X_j||)=\max\{(2\pi R)^{-d} (\angle (i,j))^{-d},1\}$.
The ``max'' operation is necessary for $\gamma$ to be a physical
``loss'' function.

Now note that two nodes will only communicate if $\angle
(i,j)<\theta_k$. The bitrate demand in Lemma \ref{prop:wk} can be
satisfied by letting $P_k=\lambda \eta_0 \ln 2 (2\pi R)^d 2^k
\theta^d_k \log^2 N$ when the bandwidths are infinite and thus the
interferences can be neglected.

Now, the probability that a node $i$ has its $k^{th}$ channel open
is at most $2^{-k}$. As such, the average power consumption of a
random node is bounded as:
\begin{eqnarray}\label{eqn:pavg}
P_{avg}&\leq& \sum_{k\in {\cal K}} 2^{-k} \times  \lambda \eta_0
\ln 2 (2\pi R)^d 2^k \theta^d_k \log^2 N\\&\leq& (\lambda \eta_0
\ln 2) (8 \pi^2)^d (\log^3 N) R^d
\end{eqnarray}

\subsection{Bandwidth Requirement}

The average power requirement in (\ref{eqn:pavg}) is found assuming
an infinite bandwidth. We now upper bound the bandwidth requirement
as a function of $N$ as follows. First note that $\forall x>0, \ln
(1+x)>x-x.^2$. Using this, Eqn. (\ref{eqn:uwbr}) can be lower
bounded as:
\begin{equation}\label{eqn:log}
R^k_t(i,j)\geq \ln 2 P^{k}\gamma(||X_i-X_j||)(\kappa-\kappa^2)
\end{equation}

where $$\kappa \triangleq \frac{1}{1+\frac{\sum_{l\in \Gamma,
l\neq j} P^{k}\phi^k_t(j,l)\gamma(||X_j-X_l||)}{\eta_0 B_k}}$$

Therefore, we only need to choose $B_k$ high enough to mask the
interference. The interference can be bounded by noticing that in
any given cap of angle $\theta_k$ there is at most one node
transmitting in channel $k$ at any given time. Now for a given
node $j$, consider the sequence of caps of angle $\theta_k,
2\theta_k, 4\theta_k,...$ centered at $j$ and call them
$C_1,C_2,C_3,...$ respectively. Now note that
$(area(C_2)-area(C_1))/area(C_1)=(1-\cos(\theta_k))/(1-\cos(\theta_k/2))-1\leq
3$, from Lemma \ref{lem:ref}. Therefore, at any given point of
time, there can be at most $3$ nodes that are in $C_2$ but not
$C_1$ and are transmitting simultaneously. Likewise, for any $m$
such that $2^m \theta_k<2\pi$, it can be shown that there are at
most $3\times 2^m$ nodes simultaneously transmitting on their
$k^{th}$ channel. The interference of these transmitting nodes on
$j$ can therefore be bounded as:
\begin{eqnarray*}
I^k&=&\sum_{l\in \Gamma, l\neq j}
P^{k}\phi^k_t(j,l)\gamma(||X_j-X_l||)\\&\leq& 6 P_k \left(8\pi R
\sqrt {2^k N^{-1} \log N}\right)^{-d}=6 P_k N^{d/2}(8 \pi)^{-d}
(2^k \log N)^{-d/2}
\end{eqnarray*}

{eqn:rk}

It therefore suffices to have $B_k>6 P_k N^{d/2}(8 \pi)^{-d} (2^k
\log N)^{-d/2}$, in which case, from (\ref{eqn:log}):
\[
R^k(i,j)\geq (\ln 2) P^{k}\gamma(||X_i-X_j||)/(4\eta_0)
\]
 Comparing with
(\ref{eqn:binf}) this indicates a factor of at most $4$ loss in
the throughout.

The total bandwidth requirement is:
\begin{eqnarray}\label{eqn:B}
B&=&\sum_{k=0}^K 6 P_k N^{d/2}(8 \pi)^{-d} (2^k \log N)^{-d/2}
\nonumber\\&=& 6 \lambda \eta_0 \ln 2 (2\pi)^d \sum_{k=0}^K
N^{d/2} (\log^{2-d/2} N) \\&\leq& 6 \lambda \eta_0 \ln 2 (2\pi)^d
N^{d/2}( \log^{3-d/2} N)
\end{eqnarray}

 We have then proved the following
Theorem:
\begin{thm}\label{main}
MRRH can provide a constant throughout of $\lambda$ with average
NL of $O(\log^3 N)$ while requiring an average power of at most
$P_{avg}=O(\lambda R^d \log^3 N )$ and a total bandwidth of at
most $B=O(\lambda N^{d/2} \log^{3-d/2} N)$. For a constant node
density $\rho$, and a direct line of sight path loss model
($d=2$), MRRH requires $P_{avg}=O(\lambda \rho^{-1} N \log^{3} N)$
and $B=O(\lambda \rho^{-1}  N \log^2 N)$.
\end{thm}

\subsection{Near Optimality of MRRH} Comparing Theorem~\ref{main} with
Theorem~\ref{prop:power} and ~\ref{prop:bnd}, the power and
bandwidth consumption of MRRH are at most $O( \log^6 N)$ away from
the absolutely most power and bandwidth aware communication
systems.

\section{Concluding Remarks}\label{sec:conc}
Low NL wireless ad-hoc networking calls for significantly more
power and bandwidth compared to nearest neighbor communication
schemes. This is because, to achieve a low NL, many long range
communications have to take place. Such communications require a
significantly larger energy to perform. These long range
communications will interfere with most other nodes in the
network. Therefore, a mere increase in the amount of the power
used in communications is not enough; the bandwidth of the system
should also increase to cancel the interferences caused by nodes
communicating over long links.

In this paper, we derived a set of new lower bound to quantify the
tradeoffs between power, bandwidth and NL. Results in propositions
\ref{prop:bnd},\ref{prop:power} put sever lower bounds on the power
and bandwidth requirement of any low NL WANET. For a close to
constant NL, one has to scale both bandwidth and average power at
least linearly with $N$.

Our bounds are tightest for small NL. By methods close to the ones
used in \cite{gupta}, one can show that even when the constraint on
NL is relaxed, a constant throughput $\lambda$ can be feasible only
if $B= \Theta(\sqrt{N})$ and $P_{avg} =\Theta(\sqrt{N})$ and is
achieved by nearest neighbor communication (NNC); note that nearest
neighbor communication incurs an average NL of of $\Theta(\sqrt{N})$
hops. Therefore, to get a factor of $\sqrt{N}$ reduction in NL
compared to NNC, one should increase the average power consumption
and bandwidth by at least a factor of $\sqrt{N}$.

Given the constraints on networks with low NL, we considered the
question of designing efficient WANETs. We devised a system, called
Multi-Resolution-Randomized-Hierarchy (MRRH) for efficient, low NL
wireless ad-hoc networking. The efficiency of MRRH was proved by
comparing its bandwidth and power requirements with our newly
derived lower bounds.

MRRH is part of our ongoing research on implications of low NL
communication on wireless ad-hoc and sensor networks
\cite{sensor}.


\bibliographystyle{unsrt}

\end{document}